\documentclass[aps,prb, twocolumn, 
groupedaddress, amsmath, superscriptaddress, showpacs, sort&compress, footinbib]{revtex4}
\usepackage{natbib}
\usepackage{colordvi}
\usepackage[dvips]{color}
\usepackage{graphicx}
\usepackage{epstopdf}
\usepackage{bm}
\usepackage{amsmath, amsfonts}
\usepackage{hyperref}
\usepackage[sort&compress]{natbib}
\begin{document}

\title{Aspects of metallic low-temperature transport in Mott-insulator/ band-insulator superlattices: optical conductivity and thermoelectricity}
\author{Andreas R\"uegg}
\affiliation{Theoretische Physik, ETH Z\"urich, CH-8093 Z\"urich, Switzerland}
\author{Sebastian Pilgram}
\affiliation{Theoretische Physik, ETH Z\"urich, CH-8093 Z\"urich, Switzerland}
\author{Manfred Sigrist}
\affiliation{Theoretische Physik, ETH Z\"urich, CH-8093 Z\"urich, Switzerland}

\date{\today}                                           % Activate to display a given date or no date

\begin{abstract}
We investigate the low-temperature electrical and thermal transport properties in atomically precise metallic heterostructures involving strongly-correlated electron systems. The model of the Mott-insulator/ band-insulator superlattice was discussed in the framework of the slave-boson mean-field approximation and transport quantities were derived by use of the Boltzmann transport equation in the relaxation-time approximation. The results for the optical conductivity are in good agreement with recently published experimental data on (LaTiO$_3)_N$/(SrTiO$_3)_M$ superlattices and allow us to estimate the values of key parameters of the model. Furthermore, predictions for the thermoelectric response were made and the dependence of the Seebeck coefficient on model parameters was studied in detail. The width of the Mott-insulating material was identified as the most relevant parameter, in particular, this parameter provides a way to optimize the thermoelectric power factor at low temperatures. 
\end{abstract}
\maketitle
\section{Introduction}
Artificial heterostructures and superlattices composed of different perovskite oxides have received a considerable attention in recent years. In many cases the heterostructures are based on two insulators that, interestingly, exhibit metallic behavior in atomically precise superlattices. For instance, Ohtomo and coworkers reported on a metallic conductivity in LaTiO$_3$/SrTiO$_3$ (Ref.~\onlinecite{Ohtomo:2002fk}) and in LaAlO$_3$/SrTiO$_3$ (Ref.~\onlinecite{Ohtomo:2004lr}) heterostructures. Thiel \emph{et al.}\cite{Thiel:2006a} demonstrated the possibility to tune the carrier density in the latter system by electric-field-effect, and Reyren \emph{et al.}\cite{Reyren08312007} detected a superconducting transition at $T_c=200$ mK. More recent studies involve LaVO$_3$/SrVO$_3$ heterostructures\cite{sheets:07} and LaVO$_3$/SrTiO$_3$ interfaces.\cite{Hotta:2007} Similar to the  LaTiO$_3$/SrTiO$_3$ system, the latter system falls into the general class of band-insulator/ Mott-insulator heterostructures.

These experimental findings have stimulated a considerable amount of theoretical work.\cite{lee:075106, Lee:2007lr, Ishida:2007, yunoki:064532, Okamoto:2004uq, okamoto:056802, Okamoto:075101, Okamoto:235108, okamoto:241104, kancharla:195427, Freericks:2006, rueegg:2007} Common to all these studies is the technical challenge to handle strong local correlations in a spatially non-uniform system and therefore calculations are mainly based on effective models. Similar to interfaces between semiconductors, electronic charges are redistributed in order to maintain an electrostatic stable solution. In the case of two insulating materials, the mutual doping can lead to metallic behavior of the interface. Generally, its electronic phase may differ from the bulk phases of the two constituents, and a variety of ordered phases in charge, spin, and orbital degrees of freedom, as well as more exotic phases, are predicted. 

In this work we are exclusively concerned with the experimentally observed metallic properties\cite{Ohtomo:2002fk, Ohtomo:2004lr, Shibuya:2004,Seo:2007}  of the interface between Mott insulator (MI) and band insulator (BI). From the theoretical point of view, there has been relatively little attempts to describe the transport properties of these quasi-two-dimensional metallic electron systems and to clarify the role of the expected strong electron-electron interaction. It is therefore of interest to study some general aspects of the electronic transport on a qualitative level. We show that experimental data\cite{Seo:2007} on the optical conductivity of LaTiO$_3$/SrTiO$_3$ superlattices were well reproduced for reasonable parameters of the present theory. In addition, we focus, in this work, on the low-temperature thermoelectricity.

\begin{figure}
\centering
\includegraphics[width=1\linewidth]{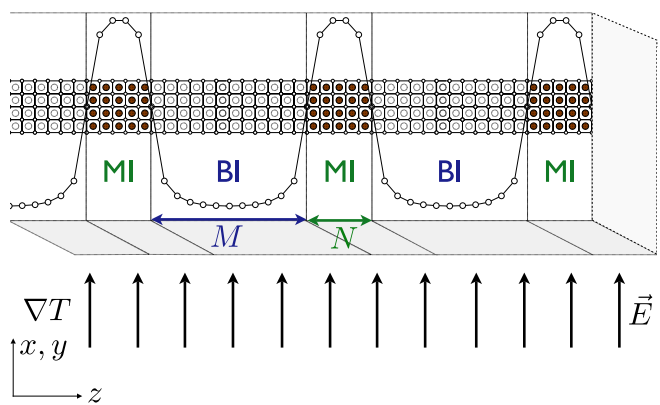}
\caption{(Color online) Schematic view of the setup considered in this article. $N$ is the number of unit cells of the Mott insulating material (MI) and $M$ is the number of unit cells of the band insulating material (BI). In addition, the conduction-band density is indicated. We assume a thermal gradient $\nabla T$ or an electrochemical field $\vec{E}$ along the $x$ direction.}
\label{fig:setup}
\end{figure}

A schematic view of the considered setup is shown in Fig.~\ref{fig:setup}. We assume a cubic lattice structure with perfect lattice match and a superlattice modulation in the $z$-direction. This setup is motivated by experiments on (LaTiO$_3)_N$/(SrTiO$_3)_M$ superlattices but other material combinations may be possible. Quite in general we assume a polar perovskite Mott insulator A$'$BO$_3$ with a $3d^1$ configuration and a nonpolar perovskite band insulator ABO$_3$ with a $3d^0$ configuration. The transport properties were calculated in response to an electrochemical field $\mathbf{E}$ or a temperature gradient $\nabla T$ in the $x$ direction. This is in contrast to previous theoretical work where the transport properties in strongly-correlated multilayer nanostructures along the $z$ direction was studied.\cite{Freericks:2006}

The motivation to study the thermoelectricity is based on several interesting observations: (i) The solid solution Sr$_{1-x}$La$_x$TiO$_3$ shows a large thermoelectric response\cite{Okuda:2001lr} for $0\leq x\leq 0.1$. (ii) From the study of narrow-band organic conductors,\cite{Chaikin:1976lr} intermetallic compounds,\cite{Miyake:2004, Zlatic:2003, milat:2006} as well as cobalt or other transition-metal oxides,\cite{Koshibae:2000fk} it is known that correlation effects can enhance thermoelectricity. (iii) Lower-dimensional structures, such as quantum-well superlattices, offer additional parameters to optimize the thermoelectric response and, at best, overtop the bulk properties.\cite{Hicks:1993lr,Ohta:2007lr}

This paper is organized as follows: In Sec. II we introduce the model. In Sec. III we discuss a quasiparticle approach to the low-energy properties and in Sec. IV the transport coefficients are calculated. A short overview of the obtained results is given in Sec. V with a subsequent discussion in Sec. VI. We summarize our main conclusions in Sec. VII.

\section{Model}
We adopt a single-band model, thereby neglecting complicating aspects regarding the orbital degrees of freedom typically present in transition metal oxides.\cite{J.Chakhalian11162007} In this model three different energy scales are important: (i) the hopping matrix element $t$, (ii) the onsite Hubbard $U$, and (iii) the Coulomb parameter of the long-range interaction $E_C=e^2/\varepsilon_Da$, involving the dielectric constant $\varepsilon_D$ of the core electron background and the lattice constant $a\approx3.9$\AA.\cite{Shibuya:2004} The onsite Hubbard $U$ is intrinsically large, namely $U>U_c$ where $U_c\approx U_c^{BR}$ is the critical interaction strength for the bandwidth controlled Mott transition in the bulk system and $U_c^{BR}\approx16t$ is the Brinkman-Rice value.\cite{Brinkman:1970lr} However, it is instructive\cite{rueegg:2007} at some points in the following discussion to vary the value of $U$ from the so-called \emph{Hartree regime} $U\ll U_c$ to the \emph{Mott regime} $U>U_c$ in order to see how the electronic properties change.\cite{comment:Uc} A realistic value of the Coulomb parameter is $E_C=0.8t$. A rough estimate of the Thomas-Fermi screening length gives $\lambda_{TF}\sim\sqrt{t/E_C}a$, which is of the order of the lattice constant $a$. In the limit $Ma\gg\lambda_{TF}$ it is, in a first approximation, sufficient to consider only one quantum well made out of $N$ layers of the MI material embedded in two semi-infinite BI systems. The microscopic model for the quantum-well system is defined by an extended single-band Hubbard Hamiltonian on a simple cubic lattice including long-range electron-electron and electron-counterion interactions as follows\cite{okamoto:241104}
\begin{eqnarray}
\hat{H}&=&-t\sum_{\langle ij\rangle\sigma}\hat{c}_{i\sigma}^{\dag}\hat{c}_{j\sigma}^{\phantom\dag}+U\sum_i\hat{n}_{i\uparrow}\hat{n}_{i\downarrow}\nonumber\\
&&+\sum_iV_i\hat{n}_i+\frac{1}{2}\sum_{i\neq j}\hat{n}_iW_{ij}\hat{n}_j.
\label{eq:mod}
\end{eqnarray}
The counterions sit in between the electronic sites, and simulate the electrostatic difference between the A (Sr$^{2+}$) and the A$'$ (La$^{3+}$) cations. The electrons at site $i$ feel an electrostatic potential
\begin{equation}
V_i=-E_{\mathrm{C}}\sum_{j}\frac{1}{|\vec{r}_i-\vec{r}_{j}^{\mathrm{ion}}|},
\end{equation}
where $\vec{r}_{j}^{\mathrm{ion}}$ denotes the position of the counterions.
The interaction matrix was given by $W_{ij}=E_C/|\vec{r}_i-\vec{r}_j|$, and we focus on charge neutral systems. We neglect effects of the lattice relaxation and fix the counterions at equidistant positions. As discussed in Ref.~\onlinecite{okamoto:056802} it was expected that the screening provided by the atomic reconstruction reduces the conduction-electron density on the central layers and enhances the density in the band insulator away from the interface. The modeling of such effects is, however, beyond the scope of this paper.

\section{Quasiparticle description}
\subsection{Single-particle Green's function}\label{sec:QP}
In the following we consider the low-temperature limit $T\rightarrow0$.
We apply a quasiparticle description for the low-energy properties of the above model system. 
Following Refs.~\onlinecite{Potthoff:1999lr} and \onlinecite{Schwieger:2003} the layer-dependent Green's function reads
\begin{equation}
G_{ll',\sigma}(\mathbf{k},\omega)=[\omega+\mu-\hat{t}(\mathbf{k})-\hat{\Sigma}_{\sigma}(\mathbf{k},\omega)]^{-1}_{ll'},
\end{equation}
where the matrix $\hat{t}(\mathbf{k})$ is the Fourier transformed hopping matrix
\begin{equation}
\hat{t}(\mathbf{k})=\left(\begin{array}{cccc}\varepsilon_{\mathbf{k}} & -t & 0& \dots\\
-t & \varepsilon_{\mathbf{k}}& -t & 0\\
0 & -t & \varepsilon_{\mathbf{k}} & -t\\
\dots &\dots&\dots&\dots
\end{array}\right),
\end{equation}
and $l,l'$ subscript the layers. The free dispersion of the two-dimensional nearest-neighbor hopping tight-binding model is $\varepsilon_{\mathbf{k}}=-2t(\cos\, k_xa +\cos\, k_ya)$ and $\hat{\Sigma}_{\sigma}(\mathbf{k},\omega)$ is the layer-dependent self-energy. We assume that the self-energy is local, thus independent of $\mathbf{k}$, and diagonal in the layer index. Furthermore, for $\omega\rightarrow 0$ we assume a paramagnetic Fermi-liquid form 
\begin{equation}
\Sigma_{ll'}(\omega)=\delta_{ll'}\left[\mu+\frac{\lambda_l}{z_l^2}+(1-\frac{1}{z_l^2})\omega\right],
\label{eq:Sigma}
\end{equation}
where we have neglected the imaginary part, which is expected to be proportional to $\omega^2$ and have suppressed the spin index $\sigma$.
From Eq.~(\ref{eq:Sigma}) it follows that the layer-dependent renormalization factor,
\begin{equation}
\hat{Z}=\left[1-\left.\frac{\partial\hat{\Sigma}(\omega)}{\partial\omega}\right|_{\omega=0}\right]^{-1},
\end{equation}
is purely local,
\begin{equation}
Z_{ll'}=\delta_{ll'}z_l^2.
\end{equation}
$\lambda_l$ in Eq.~(\ref{eq:Sigma}) is an effective layer-dependent chemical potential including Hartree-Fock like (electrostatic) contributions from the long-range Coulomb potential as well as $\omega=0$ contributions from the dynamical self-energy due to the onsite repulsion. The low-energy part of Green's function may be written in terms of an effective quasiparticle Hamiltonian
 \begin{equation}
\hat{G}(\mathbf{k},\omega)=\sqrt{\hat{Z}}\left[\omega-\hat{H}_{\mathrm{eff}}(\mathbf{k})\right]^{-1}\sqrt{\hat{Z}}+\hat{G}_{inc}(\mathbf{k},\omega),
 \end{equation}
 where the effective Hamiltonian is given by
 \begin{equation*}
\hat{H}_{\mathrm{eff}}(\mathbf{k})=\left(\begin{array}{cccc}z_1^2\varepsilon_{\mathbf{k}} +\lambda_1& -tz_1z_2 & 0& \dots\\
-tz_1z_2 & z_2^2\varepsilon_{\mathbf{k}}+\lambda_2& -tz_2z_3 & 0\\
0 & -tz_2z_3 & z_3^2\varepsilon_{\mathbf{k}} +\lambda_3& -tz_3z_4\\
\dots &\dots&\dots&\dots
\end{array}\right),
\end{equation*}
and $\hat{G}_{inc}(\mathbf{k},\omega)$ denotes the incoherent part of Green's function, which we will neglect for $\omega\rightarrow0$.
Let us introduce the eigenstates $|\mathbf{k}\nu\rangle$ of $\hat{H}_{\mathrm{eff}}(\mathbf{k})$ as
\begin{equation}
\hat{H}_{\mathrm{eff}}(\mathbf{k})|\mathbf{k}\nu\rangle=E_{\mathbf{k}\nu}|\mathbf{k}\nu\rangle,
\end{equation}
where $\nu$ is the subband index and the envelope function is given by $\psi_{\mathrm{k}\nu}(l)=\langle  l | \mathbf{k}\nu\rangle$. The $\mathbf{k}$ dependence enters via $\varepsilon_{\mathbf{k}}$,
\begin{equation}
E_{\mathbf{k}\nu}\equiv E_{\nu}(\varepsilon_{\mathbf{k}}), \quad \psi_{\mathbf{k}\nu}\equiv\psi_{\varepsilon_{\mathbf{k}}\nu}.
\end{equation}
Eventually, we find for $\omega\rightarrow0$
\begin{equation}
\left[\hat{G}(\mathbf{k},\omega)\right]_{ll'}=\sum_{\nu}\frac{z_l\psi_{\mathbf{k}\nu}(l)\psi_{\mathbf{k}\nu}(l')z_{l'}}{\omega-E_{\mathbf{k}\nu}}.
\end{equation}
Green's function for the quasiparticles is
\begin{equation}
\left[\hat{G}^{QP}(\mathbf{k},\omega)\right]_{ll'}=\sum_{\nu}\frac{\psi_{\mathbf{k}\nu}(l)\psi_{\mathbf{k}\nu}(l')}{\omega-E_{\mathbf{k}\nu}},
\end{equation}
and is diagonal in the quasiparticle subband basis.

\subsection{Slave-boson mean-field approximation}
The low-energy behavior of the self-energy may be obtained using the dynamical-mean-field-theory (DMFT) approximation\cite{okamoto:241104, Okamoto:235108, lee:075106} from which the effective quasiparticle Hamiltonian $\hat{H}_{\mathrm{eff}}(\mathbf{k})$ can be obtained. As a numerically less expensive alternative the present authors suggested in Ref.~\onlinecite{rueegg:2007} the use of a layer-dependent generalization of the Kotliar-Ruckenstein slave-boson mean-field (SBMF) approximation.\cite{Kotliar:1986kx} In the following, when it comes to actual numerical results, we rely on this approximation for the low-temperature electronic properties. (For an extended discussion on the validity of this approach see, e.g.,~Refs.~\onlinecite{Fresard:1992a} and \onlinecite{Fresard:2007fk}.) The SBMF approximation was successfully applied in other contexts to study spatially non-uniform solutions\cite{raczkowski:174525, Seibold:1998lr} and compares well with (two-site) DMFT calculations in the present context.\cite{okamoto:241104,rueegg:2007} In the SBMF approximation, one directly addresses the properties of the quasiparticles by constructing a free energy, which takes into account the competition between kinetic and potential energies. It also allows to account for local electron-correlation effects on the low-energy properties caused by the strong onsite repulsion. In particular,  $z_l$ and $\lambda_l$ are determined self-consistently in a numerically efficient way. 

The results of the SBMF approximation to the quantum-well model [Eq.~(\ref{eq:mod})] were described in detail in Ref.~\onlinecite{rueegg:2007} and we will recapitulate only some basic aspects needed for a general understanding of the present work. Within this approximation the long-range Coulomb interaction is treated in a Hartree-like way and the onsite interaction in the spirit of the Gutzwiller approximation as known from the homogeneous Hubbard model (see, e.g., Ref.~\onlinecite{Vollhardt:1984fk}). We introduce the parameter $U_r=U-E_C$. This parameter is chosen such that for $U_r=0$ the Hartree mean-field solution is recovered. We then determine self-consistently the electron-density profile in the $z$-direction, the screened effective potential $\lambda_l$, and the fraction of doubly occupied sites in each layer $l$. The self-consistent treatment  allows us to find the quasiparticle dispersion $E_{\mathbf{k}\nu}$ and the envelope function $\psi_{\mathbf{k}\nu}(l)$ of the quasiparticle $|\mathbf{k}\nu\rangle$.
Due to the strong onsite Coulomb repulsion, the quasiparticle velocity is renormalized,
\begin{equation}
\mathbf{v}_{\mathbf{k}\nu}=\frac{Z_{\mathbf{k}\nu}}{\hbar}\vec{\nabla}\varepsilon_{\mathbf{k}}.
\end{equation}
The renormalization amplitude $Z_{\mathbf{k}\nu}$ is given by
\begin{equation}
Z_{\mathbf{k}\nu}=\frac{\partial E_{\nu}(\varepsilon_{\mathbf{k}})}{\partial\varepsilon_{\mathbf{k}}}=\sum_lz_l^2\psi_{\mathbf{k}\nu}(l)^2\leq1.
\label{eq:qpv}
\end{equation}
It depends both on $\nu$ and $\varepsilon_{\mathbf{k}}$. In particular, the Fermi velocity of the subband $\nu$ is reduced by
\begin{equation}
Z_{\nu}=\left. \frac{\partial E_{\mathbf{k}\nu}}{\partial\varepsilon_{\mathbf{k}}}\right|_{E_{\mathbf{k}\nu}=E_F}.
\label{eq:mstar}
\end{equation}
For quasiparticles whose wave functions involve large contributions from the localized orbitals in the center of the Mott-insulating material, $Z_{\nu}$ is considerably smaller than one, meaning that near the chemical potential the resulting subbands are almost flat. 

\section{Calculation}
\subsection{Transport coefficient}
The self-consistent renormalized quasiparticle subbands at $T=0$ were taken to calculate the transport properties of the two-dimensional electron system. This restricts our analysis to lowest order in $T$. 

We adapt the two-dimensional version of the Boltzmann transport theory and consider the effect of the electron-electron interactions as reflected in the renormalization of the quasiparticle velocity [Eq.~(\ref{eq:qpv})] and in the $\mathbf{k}$ dependence of the envelope function $\psi_{\mathbf{k}\nu}(l)$. From a Fermi-liquid viewpoint\cite{Pines:1966} this means that we neglect the (residual) interactions between Landau's Fermi-liquid quasiparticles. The relation to the linear-response Kubo formula is discussed in Appendix~\ref{ap:1}. We introduce the local quasiparticle distribution function $f_{\mathbf{k}\nu}$. The linear response to an applied uniform electric (electrochemical) field $\mathbf{E}(\omega)$ or temperature gradient $\nabla T$ in the direction perpendicular to the growth direction of the heterostructure ($xy$-plane) is found by linearizing the distribution function $f_{\mathbf{k}\nu}=f^0(E_{\mathbf{k}\nu})+g_{\mathbf{k}\nu}$, where $f^{0}(E)=[1+\exp(\beta E)]^{-1}$ is the equilibrium Fermi-Dirac distribution function for the inverse temperature $\beta=1/k_BT$ and where $g_{\mathbf{k}\nu}$ is proportional to the applied field.

For the setup shown in Fig.~\ref{fig:setup} it is sufficient to restrict the calculations to the $x$ component of the currents. In this case we define the \emph{transport distribution function}\cite{Mahan:96} $\Phi(E)$ as follows:
\begin{equation}
\Phi(E)=\frac{e^2}{4\pi^2\hbar}\sum_{\nu}\frac{\tau_{\nu}(E)}{1-i\omega\tau_{\nu}(E)}\bar{v}_{\nu}(E)S_{\nu}(E).
\label{eq:tdf}
\end{equation}
Here, we introduced the relaxation time $\tau_{\nu}(E)$ which is assumed to be independent of the two-dimensional momentum $\mathbf{k}$. $S_{\nu}(E)$ is the area of the constant energy surface at energy $E$ of the subband $\nu$ and
\begin{equation}
\bar{v}_{\nu}(E)=\frac{1}{S_{\nu}(E)}\int|\mathbf{v}_{k\nu}|dS_{\nu}(E)
\end{equation}
is the averaged velocity over the constant energy surface. According to Eq.~(\ref{eq:tdf}) the transport distribution function is a sum of the contributions of each partially filled subband,
\begin{equation}
\Phi(E)=\sum_{\nu}\Phi_{\nu}(E).
\label{eq:Phinu}
\end{equation}
Alternatively, we introduce the layer-resolved transport distribution function as
\begin{equation}
\Phi_l(E)=\sum_{\nu}\Phi_{\nu}(E)\psi_{E\nu}(l)^2, \quad \Phi(E)=\sum_l\Phi_l(E).
\label{eq:Phil}
\end{equation}
The notation $\psi_{E\nu}(l)$ is short-hand for $\psi_{\mathbf{k}\nu}(l)$ with $E_{\mathbf{k}\nu}=E$. Equations~(\ref{eq:Phinu}) and (\ref{eq:Phil}) allow us to study the different contributions to the transport coefficients. We define the following Fermi integrals over the single kernel function $\Phi(E)$,
\begin{equation}
L^{(\alpha)}=\int dE\left(-\frac{\partial f^0}{\partial E}\right)E^{\alpha}\Phi(E).
\label{eq:FI}
\end{equation}
For the conductivity $\sigma_{2D}$, the thermopower or Seebeck-coefficient $Q_{2D}$, and the thermal conductivity $\kappa_{2D}^e$, one then finds\cite{Mahan:96, Ziman:60} 
\begin{eqnarray}
\sigma_{2D}&=&L^{(0)},\label{eq:sigma}\\
Q_{2D}&=&-\frac{1}{eT}\frac{L^{(1)}}{L^{(0)}},\label{eq:Q}
\end{eqnarray}
and
\begin{equation}
\kappa^{e}_{2D}=\frac{1}{e^2T}\left[L^{(2)}-\frac{\left(L^{(1)}\right)^2}{L^{(0)}}\right].\label{eq:kappa}
\end{equation}
Note that the thermopower $Q$ does not contain a volume term and therefore we approximate the thermopower of the superlattice as 
\begin{equation}
Q\approx Q_{2D}.
\end{equation}
On the other hand, electrical and thermal conductivities of the superlattice are approximated by
\begin{equation}
\sigma\approx\frac{\sigma_{2D}}{a(M+N)}\quad\mathrm{and}\quad\kappa^e\approx\frac{\kappa_{2D}^e}{a(M+N)}.
\end{equation}

\subsection{Impurity scattering}
\label{sec:IS}
In the following we will estimate the relaxation time $\tau_{\nu}(E)$ entering the transport distribution function [Eq.~(\ref{eq:tdf})]. We assume that at low temperatures the dominant relaxation mechanism is elastic scattering by impurities or vacancies and in the following we restrict the discussion to this case.
For simplicity we consider the short-range impurity Hamiltonian
\begin{equation}
\hat{V}_{\mathrm{imp}}=V_0\sum_{\sigma i'}\hat{c}_{i'\sigma}^{\dag}\hat{c}_{i'\sigma}^{\phantom{\dag}},
\end{equation}
where $i'$ labels the lattice sites of the impurities. We neglect multiple-scattering by different impurities and use the single-site $T$-matrix approximation\cite{Mahan:1981} with an average over the locations $i$ of the impurity. This average introduces a sum over layers $l$ and we find
\begin{equation}
\left\langle \left|T_{\nu\nu'}^{(i)}(E_{\mathbf{k}\nu})\right|^2\right\rangle_{\mathrm{av}}=\sum_l\left|\frac{\psi_{\mathbf{k}\nu}(l)V_0\psi_{\mathbf{k}'\nu'}(l)}{1-V_0G_{E_{\mathbf{k}\nu}}^{QP}(l,l)}\right|^2.
\end{equation}
Here, $T_{\nu\nu'}^{(i)}$ is the atomic $T$-matrix for a single impurity at site $i$ and
\begin{equation}
G_E^{QP}(l,l)=\sum_{\nu}\int\frac{d^2k}{(2\pi)^2}\frac{\psi_{\mathbf{k}\nu}(l)^2}{E-E_{\mathbf{k}\nu}+i0^+}
\end{equation}
is the retarded local quasiparticle Green's function of the pure system. We neglect the real part of $G^{QP}_E$ and introduce the layer-resolved quasiparticle density of states 
\begin{eqnarray}
\rho(l,E)&=&-\frac{1}{\pi}\mathrm{Im}\, G_E^{QP}(l,l)\nonumber\\
&=&\sum_{\nu}\int\frac{d^2k}{(2\pi)^2}\psi_{\mathbf{k}\nu}(l)^2\delta(E-E_{k\nu}).
\end{eqnarray}
The relaxation time follows as
\begin{equation}
\frac{1}{\tau_{\nu}(E)}=\frac{2\pi}{\hbar}V_0^2c_{\mathrm{imp}}\sum_{l}\frac{\psi_{E\nu}(l)^2\rho(l,E)}{1+\pi^2V_0^2\rho(l,E)^2},
\label{eq:taunu}
\end{equation}
where $c_{\mathrm{imp}}$ is the impurity concentration assumed to be independent of the layer $l$.

We note that even though the energy dependence of the relaxation time near the Fermi energy is rather weak, varying the potential strength $V_0$ has a significant influence on the transport properties and, in particular, on the scaling behavior of these quantities with $N$ (see Sec.~\ref{Results:FCR}). The structure of Eq.~(\ref{eq:taunu}) offers the possibility to smoothly vary the relaxation time in each subband and to study different scenarios of quasiparticle scattering. Due to the spatial non-uniformity, the value of $\rho(l,0)$ strongly changes with $l$ reaching its maximum in the center of the heterostructure where the quasiparticles states are confined to energies near the Fermi energy $E=0$.  The first Born approximation is recovered in the limit of weak impurities: $\pi V_0\rho(l,0)\ll1$ for all $l$. In this case, $\tau_{\nu}(0)$ is substantially smaller for the subbands whose spatial weight is mostly located in the center of the quantum well. In other words, slower quasiparticles are scattered more strongly. In this case, the quasiparticle transport is mainly determined by the properties of the almost empty subbands. In the unitary limit, $\pi V_0\rho(l,0)\gg1$, the scattering rate of the slower quasiparticles was significantly reduced. In this case, the quasiparticle transport is mainly determined by the strongly renormalized subbands. For moderate values of $V_0$, the dependence of the scattering rate on the subband index was less pronounced.

Beside the impurity potential strength, there are other parameters that can influence the subband dependence of $\tau_{\nu}$ in real systems. For example, it is conceivable that the scattering occurs mainly due to imperfections of the interface. Furthermore, there is no reason not to believe that at finite temperatures electron-electron or electron-phonon scattering will affect the individual subbands in a different way. In this sense, Eq.~(\ref{eq:taunu}) can be considered to be a convenient form to vary the subband dependence by a single parameter.

\section{Overview}
\subsection{Optical conductivity and free carrier response}
\begin{figure}
\includegraphics[width=0.8\linewidth]{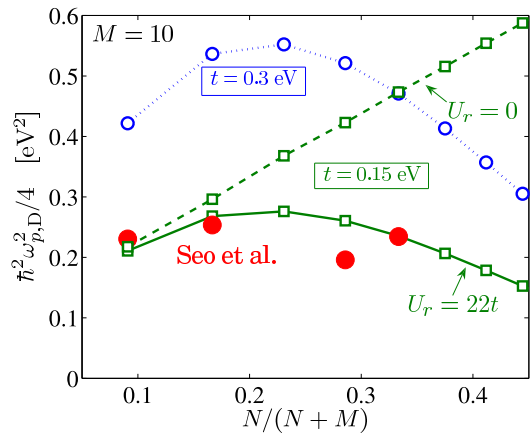}
\caption{(Color online) The plasma frequency $\omega_{pD}$ of (MI)$_N$/(BI)$_{M=10}$ superlattices, as found by the present theory for $t=0.3$ eV, and $U_r=22t$ (dotted line), $t=0.15$ eV and $U_r=0$ (dashed line) and $t=0.15$ eV and $U_r=22t$ (solid line). The filled circles denote the measurements of Seo et al. on (LaTiO$_3)_N$/(SrTiO$_3)_{M=10}$ superlattices reported in Ref.~\onlinecite{Seo:2007}.} 
\label{fig:drude}
\end{figure}

Figure~\ref{fig:drude} shows a comparison of the calculated oscillator strength $\omega_{pD}$ of the Drude peak in the optical conductivity with the results of optical experiments reported by Seo et al. in Ref.~\onlinecite{Seo:2007} on (LaTiO$_3)_N$/(SrTiO$_3)_{M=10}$ superlattices. There is a good qualitative agreement between the prediction from the theory in the Mott regime $U_r>U_c^{BR}\approx 16t$ and the measurements. Even a rather good quantitative agreement is achieved by choosing the hopping matrix element of the order $t=0.15$ eV. In contrast, in the Hartree regime $U_r\ll U_c^{BR}$, the qualitative behavior of $\omega_{pD}$ as function of $N/(N+M)$ is quite different and clearly not consistent with the experimental data.

In Sec.~\ref{Results:FCR} we also discuss the possibility to gain information about the scattering parameters by analyzing the width of the measured Drude peak.

\subsection{Thermoelectricity}
\begin{figure}
\includegraphics[width=0.8\linewidth]{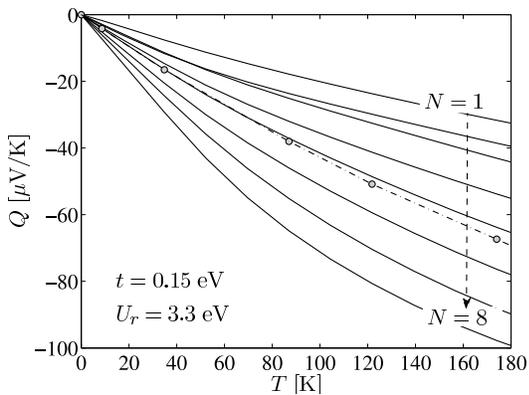}
\caption{The temperature dependence of the thermopower $Q$ for quantum wells with $N=1,\dots, 8$, using the first Born approximation for the scattering process. The dotted-dashed line indicates the result for the $N=5$ heterostructure when taking the self-consistent renormalized subbands at finite temperatures. The values of the relevant parameters are indicated in the figure.}
\label{fig:QT}
\end{figure}
\begin{figure}
\includegraphics[width=0.8\linewidth]{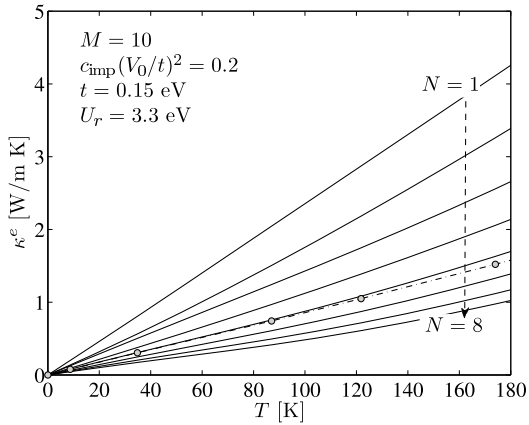}
\caption{The electronic contribution $\kappa^e$ to the total heat conductivity for the same parameters as in Fig.~\ref{fig:QT}. The dotted-dashed line indicates the result for the $N=5$ heterostructure when taking the self-consistent renormalized subbands at finite temperatures.}
\label{fig:kappaN}
\end{figure}
\begin{figure}
\includegraphics[width=0.8\linewidth]{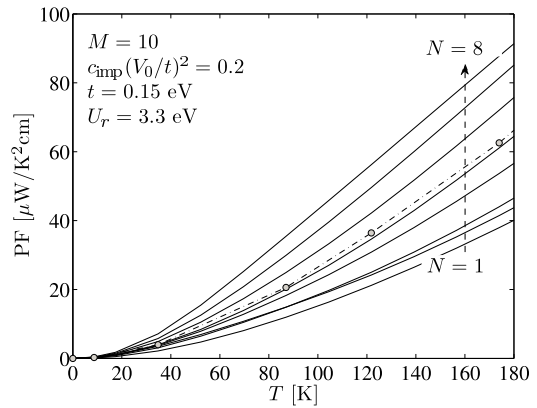}
\caption{The power factor PF=$\sigma Q^2$ for the same parameters as in Fig.~\ref{fig:kappaN}. The dotted-dashed line has the same meaning as in Fig.~\ref{fig:kappaN}.}
\label{fig:PF}
\end{figure}
Due to the uncertainty in the subband dependence of the relaxation time $\tau_{\nu}$ it is difficult to estimate the thermoelectric transport even at low temperatures. As an illustrative example we compare the results using Eq.~(\ref{eq:taunu}) in the first Born approximation and in the unitary limit. 

We first show the results of the Born approximation. Figures \ref{fig:QT}, \ref{fig:kappaN} and \ref{fig:PF} show the calculated low-temperature thermoelectric power (Seebeck coefficient) $Q$, the electronic contribution $\kappa^e$ to the thermal conductivity, and the power factor PF$=\sigma Q^2$ as function of temperature $T$ for different width $N$ of the MI region, respectively. All the calculated transport quantities show the typical metallic low-temperature dependence.  In particular, the thermopower $Q$ is negative and its magnitude increases linearly with $T$. The numerical values of the relevant parameters were estimated by comparing the theoretical results to transport\cite{Shibuya:2004} and spectroscopic\cite{Seo:2007} measurements of SrTiO$_3$/LaTiO$_3$ superlattices and are indicated in Figs.~\ref{fig:QT} -  \ref{fig:PF}. In order to check the validity of the zero-temperature approximation for the quasiparticle subbands, we also show the results for the $N=5$ heterostructure as dotted-dashed line when taking the self-consistent renormalized subbands at finite temperatures. The difference between the two results is of order $T^2$ and in the considered temperature range, it is a good approximation to use the zero-temperature subbands for the calculations.

\begin{figure}
\includegraphics[width=0.9\linewidth]{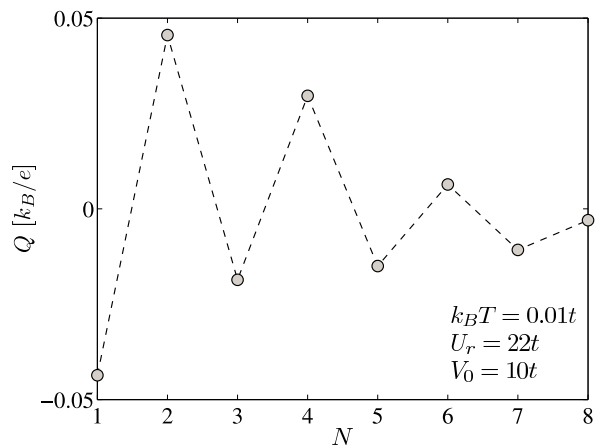}
\caption{The thermopower $Q$ at $k_BT=0.01t$ for $N=1-8$ in the limit of a strong impurity potential $V_0=10t$.}
\label{fig:QNV10}
\end{figure}

In the unitary limit, a completely different behavior of the low-temperature thermopower is observed. In Fig.~\ref{fig:QNV10} the Seebeck coefficient is shown for different width $N$ of the quantum well. Interestingly, the value of the low-temperature thermopower $Q$ shows an even-odd oscillation with a decreasing amplitude for $N\rightarrow\infty$.

These results show that the Seebeck coefficient is in fact a sensitive probe of the scattering mechanism. Measurements of this quantity could reveal important informations about which subbands contribute dominantly to the transport. We present a more detailed analysis on this point in the next section.

\section{\label{Results:FCR}Discussion}
\subsection{Optical conductivity}

The optical conductivity is considered in the regime $\omega\tau_{\nu}\gg~1$ of Eq.~(\ref{eq:tdf}). In this regime the conductivity $\sigma_{2D}(\omega)$ at low temperatures has the well-known Drude form:
\begin{equation}
\sigma_{2D}(\omega)=\frac{D}{\pi}\frac{\bar{\tau}}{1-i\omega\bar{\tau}}.
\label{eq:sigmacl}
\end{equation}
On the right-hand side of Eq.~(\ref{eq:sigmacl}) we introduced the (two-dimensional) Drude weight $D=\sum_{\nu}D_{\nu}$. The contributions of the partially filled subbands are
\begin{eqnarray}
D_{\nu}&=&\frac{e^2}{4\pi\hbar}S_{\nu}(0)\bar{v}_{\nu}(0)\nonumber\\
&=&e^2\pi\frac{Z_{\nu}\mathcal{N}_v(\varepsilon_{\nu}^*)}{\hbar^2}.
\label{eq:Drude}
\end{eqnarray}
Here, $\varepsilon_{\nu}^*$ is defined by the equation $E_{\nu}(\varepsilon_{\nu}^*)=0$ and 
\begin{equation}
\mathcal{N}_v(\varepsilon)=\int\frac{d^2k}{(2\pi)^2}|\nabla\varepsilon_k|^2\delta(\varepsilon-\varepsilon_k)
\end{equation}
is the weighted density of states.\cite{comment:drude} The width $1/\bar{\tau}$ of the Drude peak is obtained by the following weighted sum
\begin{equation}
\frac{1}{\bar{\tau}}=\sum_{\nu}\frac{D_{\nu}}{D}\frac{1}{\tau_{\nu}},
\label{eq:taubar}
\end{equation}
where $\tau_{\nu}$ is the relaxation time of the subband $\nu$. 

\begin{figure}
\includegraphics[width=0.9\linewidth]{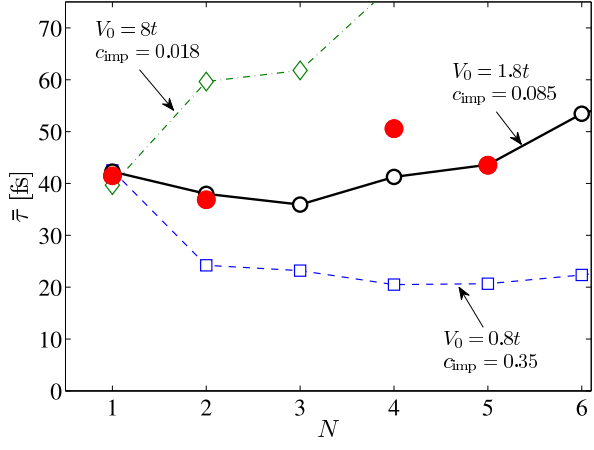}
\caption{(Color online) The averaged relaxation time $\bar{\tau}$ as found in the present theory ($t=0.15$ eV) and $U_r=22t$ for different scattering parameters as function of $N$.  The filled circles denote the inverse width of the measured Drude peak in (LaTiO$_3)_N$/(SrTiO$_3)_{M=10}$ superlattices reported in Ref.~\onlinecite{Seo:2007}.} 
\label{fig:taubar}
\end{figure}

Experimentally, the Drude weight and the plasma frequency $\omega_{pD}$ are usually determined by fitting the complex dielectric function to a Drude+Lorentz form.\cite{Seo:2007} In order to compare the calculated total Drude weight $D=\sum_{\nu}D_{\nu}$ of the two-dimensional electron system with the plasma frequency $\omega_{pD}$ measured in optical experiments on (MI)$_N$/(BI)$_M$ superlattices we assume independent quantum wells that yields the relation  
\begin{equation}
\frac{\omega_{pD}^2}{4}=\frac{D}{aN}\frac{N}{N+M}.
\label{eq:omegapD}
\end{equation}
In this expression $D=D(N,U,E_C,t)$ is a function of the quantum well parameters. The calculations of $\omega_{pD}$ using Eq.~(\ref{eq:omegapD}) are shown in Fig.~\ref{fig:drude}. In order to gain information about the scattering parameters, we assume that all the samples are of equal quality, $c_{\mathrm{imp}}(N)=\mathrm{const.}$, and compare in Fig.~\ref{fig:taubar} the value of $\bar{\tau}$ as a function of $N$ with the width of the measured Drude peak. For a moderate impurity strength $V_0\approx 1.8t$ and a reasonable impurity concentration $c_{\mathrm{imp}}\approx 0.085$, the scaling of $\bar{\tau}$ with $N$, as well as its absolute value, is in good agreement with the experimental results. For these parameters the mean relaxation time is $\bar{\tau}\approx 40$ fs.

\subsection{DC conductivity}
The electrical and thermal conductivities were obtained in the static limit $\omega\tau_{\nu}\ll1$ of Eq.~(\ref{eq:tdf}). From Eq.~(\ref{eq:sigma}) it follows that at low temperatures the dc conductivity is constant and given by
\begin{equation}
\sigma_{2D}=\Phi(0)=\frac{e^2}{4\pi^2\hbar}S_FL.
\end{equation}
$S_F=\sum_{\nu}S_{\nu}(0)$ is the total area of the Fermi surface and the mean free path is obtained as a weighted average,
\begin{equation}
L=\frac{1}{S_F}\sum_{\nu}S_{\nu}(0)\bar{v}_{\nu}(0)\tau_{\nu}(0).
\end{equation}
With the parameters estimated in the last section for the (LaTiO$_3)_N$/(SrTiO$_3)_{M=10}$ superlattices,
we find an averaged mean-free path of $L\approx 10\, a$ which is about half of the value given in Ref.~\onlinecite{Seo:2007}. 

The electronic contribution to the thermal conductivity follows from Eq.~(\ref{eq:kappa}), 
\begin{equation}
\kappa^e_{2D}=\frac{\pi^2}{3}\frac{k_B^2T}{e^2}\Phi(0),
\end{equation}
and is related to the electrical conductivity by the Wiedemann-Franz law typical for elastic scattering.\cite{ashcroftmermin} 

\subsection{Thermopower or Seebeck coefficient}
Next we consider the thermopower $Q$, which, in an open circuit, is defined as the constant between electrochemical field and temperature gradient $\mathbf{E}=Q\nabla T$. For a metal at low temperatures, it follows from Eq.~(\ref{eq:Q}) that $Q$ is given by\cite{ashcroftmermin}
\begin{equation}
Q=-\frac{\pi^2}{3}\frac{k_B^2T}{e}\left.\frac{\partial}{\partial E}\log\Phi(E)\right|_{E=0}.
\end{equation}

In the quantum-well system, we can interpret the thermopower as a weighted sum of the contributions of the partially filled quasiparticle subbands, 
\begin{equation}
Q=\frac{\sum_{\nu}Q_{\nu}\sigma_{\nu}}{\sigma},
\label{eq:Qtotnu}
\end{equation}
where $\sigma_{\nu}$ and $Q_{\nu}$ are the conductivity and the thermopower of the subband $\nu$, respectively. Alternatively, one considers a weighted sum of the individual layers,
\begin{equation}
Q=\frac{\sum_lQ_l\sigma_l}{\sigma}.
\label{eq:Qtotl}
\end{equation}
In the quasiparticle description, it is more natural to use the first interpretation and, in the following, we will focus on the multi-subband aspect.

When a rigid-band picture is applicable, meaning that the dispersion $E_{\mathbf{k}\nu}$ is independent of the filling of the band, one can show that a multi-band system yields a small magnitude of $Q$. It is therefore less efficient compared to the single-band system with the best properties because of possible cancellations in the enumerator of Eq.~(\ref{eq:Qtotnu}) due to electron-like ($Q_{\nu}<0$) and hole-like ($Q_{\nu}>0$) contributions [see, e.g., Ref.~\onlinecite{Hicks:1993lr}]. However, this conclusion breaks down in strongly-correlated electron systems because the rigid-band picture is in general not applicable and therefore we believe that a multi-band system is not a priori less efficient. In fact, the effect of onsite correlations is strongest near half filling which leads to several overlapping subbands in superlattice structures.

\subsubsection{Thermopower: subband contributions}
The value of the thermopower for an individual subband is
\begin{equation}
Q_{\nu}=-\frac{\pi^2}{3}\frac{k_B}{e}\frac{k_BT}{Z_{\nu}}\left[\frac{\tau_{\nu}'(\varepsilon_{\nu}^*)}{\tau_{\nu}(\varepsilon_{\nu}^*)}+\frac{\mathcal{N}_v'(\varepsilon_{\nu}^*)}{\mathcal{N}_v(\varepsilon_{\nu}^*)}+\frac{Z_{\nu}'}{Z_{\nu}}\right],
\label{eq:Qnu}
\end{equation}
where $\tau_{\nu}(\varepsilon)\equiv\tau_{\nu}(E_{\nu}(\varepsilon))$ and the prime denotes the derivative with respect to $\varepsilon$. There is an overall reduction of the energy scale by $Z_{\nu}$. As a result, the low-temperature slope of $Q_{\nu}$ is enhanced by a factor of $1/Z_{\nu}$ due to correlation effects.\cite{comment:Znu} In the following we will discuss the different contributions to $Q_{\nu}$.

The first term inside the square brackets of Eq.~(\ref{eq:Qnu}) describes the influence of the scattering process. Usually, when considering elastic impurity scattering, this term is small and can be neglected. The second term describes the contribution from the uncorrelated band structure. It is sizable if the subband occupation $n_{\nu}$ is small, because the particle-hole asymmetry is large near the subband edge. In fact, for the almost empty subbands with $n_{\nu}a^2\ll1$, we find 
\begin{equation}
\frac{\mathcal{N}_v'(\varepsilon_{\nu}^*)}{\mathcal{N}_v(\varepsilon_{\nu}^*)}\approx\frac{1}{4t+\varepsilon_{\nu}^*}=\frac{1}{2t\pi n_{\nu}a^2}.
\label{eq:Np}
\end{equation}
On the other hand, in the presence of strong electron correlations, the third term $Z_{\nu}'/Z_{\nu}$ is dominant for most subbands because the spatial non-uniformity of the system leads to a sizable logarithmic derivative of the quasiparticle weight at the Fermi energy.\cite{rueegg:2007, comment:hf}

\begin{figure}
\centering
\includegraphics[width=1\linewidth]{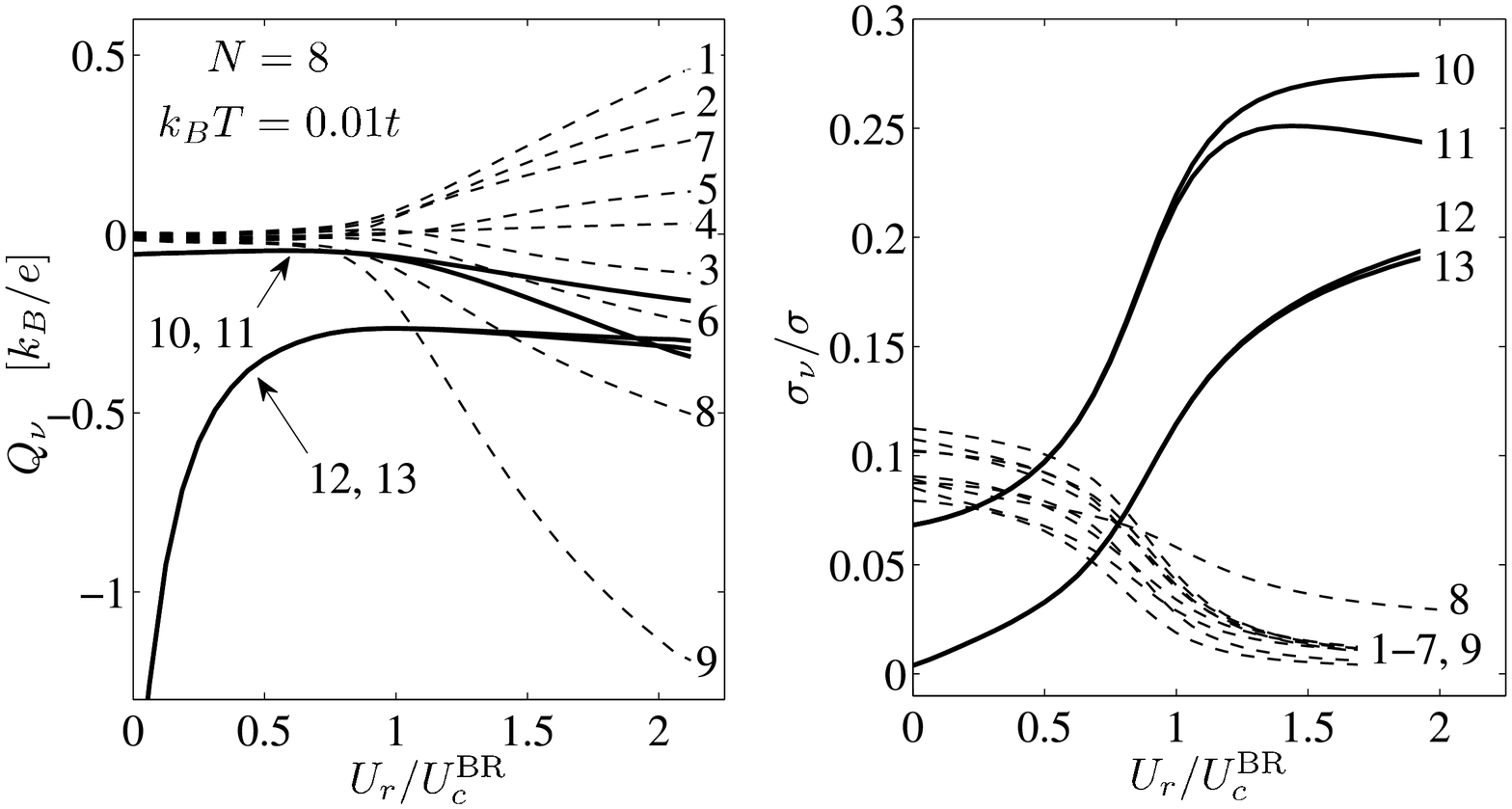}
\caption{The thermopower $Q_{\nu}$ and the relative conductivity $\sigma_{\nu}/\sigma$ of the individual subbands as function of $U_r/U_c^{\mathrm{BR}}$ for the $N=8$ quantum well. The numbers on the right-hand side indicate the subband index in descending order in the subband occupation starting with $\nu=1$ with the highest subband occupation. }
\label{fig:Qnusigmanu}
\end{figure}

The dominance of the third term becomes evident in a study of the thermopower as onsite correlation effects are varied. $Q_{\nu}$ is shown in the left panel of Fig.~(\ref{fig:Qnusigmanu}) as function of $U_r/U_c^{\mathrm{BR}}$ for the $N=8$ heterostructure at $k_BT=0.01t$. For the strongly renormalized subbands ($\nu=1-9$ in Fig.~\ref{fig:Qnusigmanu}), we observe a clear qualitative difference between Hartree ($U_r\ll U_c^{\mathrm{BR}}$) and Mott ($U_r>U_c^{\mathrm{BR}}$) regimes. In particular, in the Mott regime, there is a noticeable spread of $Q_{\nu}$ mainly due to the third term in Eq.~(\ref{eq:Qnu}). On the other hand, the thermopower of the weakly renormalized subbands ($\nu=10-13$ in Fig.~\ref{fig:Qnusigmanu}) is governed by Eq.~(\ref{eq:Np}) and the change as function of $U_r/U_c^{\mathrm{BR}}$ is due to the change of the subband occupation $n_{\nu}$. Since $\tau_{\nu}'/\tau_{\nu}\ll1$, the qualitative behavior of the \emph{subband} thermopower $Q_{\nu}$ is almost independent of the value of the impurity strength $V_0$.

\subsubsection{Dependence of the total thermopower on the impurity potential}
It is crucial for the \emph{total} thermopower how the different contributions $Q_{\nu}$ are weighted. 
According to Eq.~(\ref{eq:Qtotnu}) the weighting factor is $\sigma_{\nu}/\sigma$, which depends on the Drude weight $D_{\nu}$ and on the relaxation time $\tau_{\nu}$. As an example, $\sigma_{\nu}/\sigma$ is shown in the right panel of Fig.~\ref{fig:Qnusigmanu} using the Born approximation of Eq.~(\ref{eq:taunu}) corresponding to $V_0\rightarrow0$. In this case, the most dominant contributions to $Q$ in the Mott regime arise from the highest partially filled subbands ($\nu=10-13$ in Fig.~\ref{fig:Qnusigmanu}). The reason is twofold: first, electron-like and hole-like contributions of the correlated subbands partially cancel, and second, the relative conductivity $\sigma_{\nu}/\sigma$ of these bands is significantly reduced.

\begin{figure}
\centering
\includegraphics[width=0.8\linewidth]{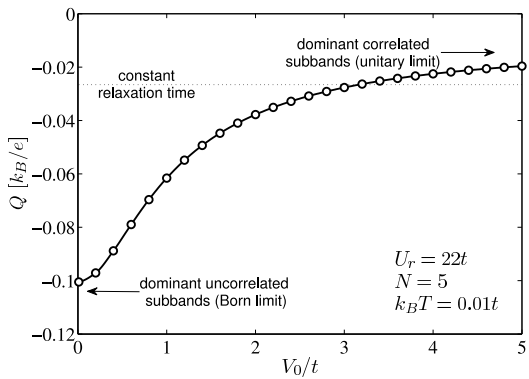}
\caption{The thermopower $Q$ of the $N=5$ heterostructure as function of the impurity potential $V_0$, which governs the relative contributions of the individual subbands. Also indicated is the value of $Q$ when assuming a constant relaxation time.}
\label{fig:QN5V}
\end{figure}

However, for larger values of $V_0$, the contributions of the subbands subject to strong correlation effects become more and more important. Figure~\ref{fig:QN5V} shows the dependence of $Q$ on the potential $V_0$ for $N=5$. For $V_0\rightarrow0$ the weakly renormalized subbands are dominant, leading to a relatively large negative thermopower. For $V_0\rightarrow\infty$ the thermopower increases because the positive contributions of some of the correlated bands gain more weight (see also Fig.~\ref{fig:Qnusigmanu}). Interestingly, for an even $N$, we observe that $Q$ can take positive values at low temperatures because there is one subband close to half filling with a dominant positive contribution. As a consequence, $Q$ oscillates as function of $N$ (see Fig.~\ref{fig:QNV10}). Moreover, the larger the $N$ the larger the effect of cancellation becomes, which reduces the amplitude of the oscillation for $N\rightarrow\infty$.

\subsection{Thermoelectric application}
For thermoelectric applications, the dimensionless figure of merit $Z_TT=\sigma Q^2T/(\kappa^e+\kappa^L)$ should be as large as possible. This quantity provides a measure of the efficiency of a material used for cooling or heating, respectively, as a thermoelectric power generator. It involves the static transport coefficients of the system, the electrical conductivity $\sigma$, the thermoelectric power or Seebeck coefficient $Q$ and the total thermal conductivity $\kappa=\kappa^e+\kappa^L$. A high value of $\sigma$ is necessary to minimize Joule heating, while a low value of $\kappa$ helps to maintain a large temperature gradient. In metal-oxide based thermoelectric materials, the phonon thermal conductivity usually plays a dominant role and often $\kappa^L\gg\kappa^e$ (see, e.g., Ref.~\onlinecite{Tokura:1993lr} and \onlinecite{Ohta:2007lr}). Therefore, we will focus on a quantity involving only the electronic transport properties, namely the product PF$=\sigma Q^2$ called the power factor. It is a common approach to optimize PF by varying external parameters, and we will discuss the possibility to optimize PF by changing the superlattice parameters $N$ and $M$.\cite{Hicks:1993lr,Broido:1995lr}

\begin{figure}
\includegraphics[width=0.8\linewidth]{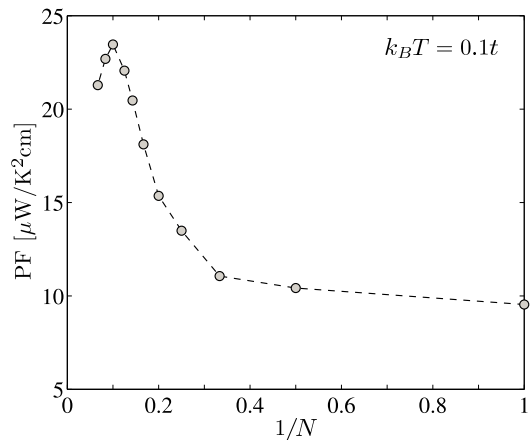}
\caption{The thermopower PF$=\sigma Q^2$ as function of the inverse width $1/N$ of the Mott-insulating material at $k_BT=0.1t$. The chosen parameters are as follows: the onsite interaction is $U_r=22t$, the width of the band-insulator is $M=10$, and the impurity scattering in the first Born approximation is parameterized by $(V_0/t)^2c_{\mathrm{imp}}=0.2$.}
\label{fig:PFN}
\end{figure}

Let us, for example, fix $M=10$. Since both $Q$ and $\sigma$ depend also on $N$ one can try to optimize the power factor by varying $N$. For the case of weak impurities 
this is shown in Fig.~\ref{fig:PFN} for a temperature $k_BT=0.1t$. We find that for this temperature the power factor takes the largest value at $N\sim10$. The maximum, as function of $N$, arises due to the competition between an enhancement of $Q^2$ by increasing $N$ (see Fig.~\ref{fig:QT}) and the volume term $\sim 1/(N+M)$, which reduces the total conductivity $\sigma$. 

\section{Conclusions}

In this work we have focused on the metallic properties observed in Mott-insulator/ band-insulator heterostructures. We propose an analysis based on a quasiparticle picture that is expected to be valid in the Fermi liquid regime below the coherence temperature. A self-consistent renormalization of the quasiparticles was obtained by applying the slave-boson mean-field approximation and the quasiparticle transport was calculated in the relaxation-time approximation of the Boltzmann transport equation.

The calculated optical conductivity is in good agreement with previously published spectroscopic measurements. Furthermore, the comparison with experiment allows to extract some of the key parameters of the model.

We have shown that the correlation effects enhance the thermoelectric response as compared to a metal/ band-insulator interface. Furthermore, we have demonstrated that the design of the superlattice offers in principle experimentally accessible parameters to optimize the power factor for a given temperature. However, the thermoelectric response, in particular the Seebeck coefficient, is very sensitive to the considered scattering mechanism. Even when focusing on $s$-wave impurity scattering at low temperatures, two completely different low-temperature behavior of $Q$ can be obtained in the Born and the unitary limit. Especially, this difference is manifest when considering $Q$ as function of the width $N$ of the quantum well. On the other hand, measurements of the thermopower would actually be a good probe to test to which extent the quasiparticle description holds and, where applicable, to gain information on the delicate point of which subband contributions are most dominant for the quasiparticle transport.

Whether the thermoelectric properties of a superlattice of the kind considered in this paper can overtop those of a bulk solid solution with the same composite remains unclear and further studies are necessary to clarify this point. In particular, we have neglected aspects concerned with the orbital and spin degrees of freedom, which may lead to ordered phases modifying the simple picture given here. Furthermore, our results are limited to the Fermi-liquid regime with coherent quasiparticle transport. It is possible that the qualitative behavior at higher temperatures is quite different. For example, it is known that the thermopower in the single-band Hubbard model may even change sign with increasing temperature indicating the transition from electron-like coherent to hole-like incoherent transport.\cite{Chaikin:1976lr,Palsson:1998lr} Similar effects may also occur in the model discussed in this paper. Furthermore, for higher temperatures finite-lifetime aspects due to the electron-electron interactions and due to the coupling to the lattice degrees of freedom play a cumulative role. In this case a more sophisticated analysis is necessary.

\begin{acknowledgments}
We would like to thank K.~Miyake, S.~S.~A.~Seo, N.~Kawakami, V.~Zlatic, R.~Asahi, H.~Adachi, and I.~Milat for valuable discussions. This study was financially supported by the Toyota Central R \& D Laboratories, Nagakute, Japan and the NCCR MaNEP of the Swiss Nationalfonds.
\end{acknowledgments}
\begin{appendix}
\section{}\label{ap:1}
In order to clarify the approximations undertaken in the Boltzmann description for the dc transport coefficients, we derive the transport distribution function $\Phi(E)$ starting from the linear-response Kubo formula\cite{Mahan:1981} and assuming the validity of the quasiparticle description outlined in Sec.~\ref{sec:QP}. This requires the evaluation of the current-current correlation function
\begin{equation}
\pi(i\omega)=-\frac{1}{2\Omega}\int_0^{\beta}e^{i\omega\tau}\left\langle T_{\tau}\vec{j}(\tau)\cdot\vec{j}(0)\right\rangle,
\label{eq:cc}
\end{equation}
where $\Omega=N_{||}a^2$ is the sheet volume. A subsequent analytical continuation $i\omega\rightarrow\omega+i\delta$ yields the sheet dc conductivity by taking the following limit
\begin{equation}
\sigma_{2D}=-\lim_{\omega\rightarrow0}\left[\frac{\mathrm{Im}\, \pi^{R}(\omega)}{\omega}\right].
\end{equation}
The remaining transport coefficients may be obtained by applying the Jonson-Mahan theorem,\cite{Jonson:1980lr} which holds for the model [Eq.~(\ref{eq:mod})] when replacing the long-range interaction part by a (self-consistently determined) one-particle potential.\cite{Freericks:2007fk} The current operator for this model and for transport in the $x, y$ direction is ($\hbar=1$)
\begin{equation}
\vec{j}=-e\sum_{l\mathbf{k}\sigma}\nabla\varepsilon_{\mathbf{k}}c_{l\mathbf{k}\sigma}^{\dag}c_{l\mathbf{k}\sigma}^{\phantom{\dag}},
\end{equation}
$\mathbf{k}$ is a two-dimensional crystal momentum and $l$ the layer index. In the following we show that when neglecting vertex corrections in Eq.~(\ref{eq:cc}), one arrives at the transport distribution function [Eq.~(\ref{eq:tdf})]. Thus, we approximate
\begin{equation*}
\pi(i\omega)=\frac{e^2}{\Omega\beta}\sum_{\mathbf{k}}\left|\nabla\varepsilon_{\mathbf{k}}\right|^2\sum_{ip_n}\mathrm{Tr}\left[\hat{\mathcal{G}}(\mathbf{k},ip+i\omega)\hat{\mathcal{G}}(\mathbf{k},ip)\right],
\end{equation*}
where the trace has to be performed over the different layers and $\mathcal{G}$ is the full Matsubara Green's function including all self-energy corrections. After letting $i\omega\rightarrow\omega+i\delta$, one finds the form given in Eq.~(\ref{eq:sigma}),
\begin{equation}
\sigma_{2D}=\int dE\left(-\frac{\partial f}{\partial E}\right)\Phi(E)
\end{equation}
with the transport distribution function
\begin{equation}
\Phi(E)=\frac{\pi e^2}{\Omega}\sum_{\mathbf{k}}\left|\nabla\varepsilon_{\mathbf{k}}\right|^2\mathrm{Tr}\left[\hat{A}(\mathbf{k},E)^2\right].
\label{eq:PhiA}
\end{equation}
Here, we have introduced the spectral density matrix $\hat{A}(\mathbf{k},E)=-\frac{1}{\pi}\mathrm{Im}\, \hat{G}^R(\mathbf{k},E)$ for which we assume (see Sec.~\ref{sec:QP}) the following low-energy form
\begin{equation}
\left[\hat{A}(\mathbf{k},E)\right]_{ll'}=\frac{1}{\pi}\sum_{\nu}\frac{z_l\psi_{\mathbf{k}\nu}(l)\gamma_{\mathbf{k}\nu}\psi_{\mathbf{k}\nu}(l')z_{l'}}{(E-E_{\mathbf{k}\nu})^2+\gamma_{\mathbf{k}\nu}^2}.
\end{equation}
The quasiparticle life-time is given by the relation $\gamma_{\mathbf{k}\nu}=1/2\tau_{\nu}(E_{\mathbf{k}\nu})$, where 
\begin{equation}
\gamma_{\mathbf{k}\nu}=-\sum_lz_l^2\psi_{\mathbf{k}\nu}(l)^2\left[\mathrm{Im}\, \hat{\Sigma}(E_{\mathrm{k}\nu})\right]_{ll}.
\end{equation}
In lowest order in temperature we can restrict the analysis to the Fermi surface. In this case $\gamma_{\mathbf{k}\nu}$ is proportional to the impurity concentration $c_{\mathrm{imp}}$ for dilute impurities. For non-degenerate subbands and in the limit $c_{\mathrm{imp}}\rightarrow 0$ only terms, which are diagonal in the subband index, contribute to the trace in Eq.~(\ref{eq:PhiA}). After using the relation [Eq.~(\ref{eq:qpv})] for the renormalization amplitude $Z_{\mathbf{k}\nu}$ one arrives at Eq.~(\ref{eq:tdf}) obtained from the linearized Boltzmann equation (for $\omega=0$). In actual calculations we estimate $\gamma_{\mathbf{k}\nu}$ by the approximations discussed in Sec.~\ref{sec:IS}.

In summary, we assume the validity of a Fermi-liquid description with local self-energy corrections, dominant $s$-wave scattering on dilute impurities, and that effects of weak localization can be neglected. Under these conditions we expect that, referring to the homogenous and isotropic system,\cite{Betbeder:1966} the transport distribution function used in the present work captures the main features in lowest order in temperature.

\end{appendix}
\bibliography{references}
\end{document}